\input harvmac
\let\includefigures=\iftrue
\let\useblackboard=\iftrue
\newfam\black

\includefigures
\message{If you do not have epsf.tex (to include figures),}
\message{change the option at the top of the tex file.}
\input epsf
\def\figin{\epsfcheck\figin}\def\figins{\epsfcheck\figins}
\def\epsfcheck{\ifx\epsfbox\UnDeFiNeD
\message{(NO epsf.tex, FIGURES WILL BE IGNORED)}
\gdef\figin##1{\vskip2in}\gdef\figins##1{\hskip.5in}
\else\message{(FIGURES WILL BE INCLUDED)}%
\gdef\figin##1{##1}\gdef\figins##1{##1}\fi}
\def\DefWarn#1{}
\def\figinsert{\goodbreak\midinsert}
\def\ifig#1#2#3{\DefWarn#1\xdef#1{fig.~\the\figno}
\writedef{#1\leftbracket fig.\noexpand~\the\figno}%
\figinsert\figin{\centerline{#3}}\medskip\centerline{\vbox{
\baselineskip12pt\advance\hsize by -1truein
\noindent\footnotefont{\bf Fig.~\the\figno:} #2}}
\endinsert\global\advance\figno by1}
\else
\def\ifig#1#2#3{\xdef#1{fig.~\the\figno}
\writedef{#1\leftbracket fig.\noexpand~\the\figno}%
\global\advance\figno by1} \fi

\def\id{{1 \kern-.28em {\rm l}}}

\def\K3{{\bf K3}}
\def\journal#1&#2(#3){\unskip, \sl #1\ \bf #2 \rm(19#3) }
\def\andjournal#1&#2(#3){\sl #1~\bf #2 \rm (19#3) }

\def\bar{\overline}

\def\tilde{\widetilde}

\def\frac#1#2{{#1\over#2}}

\def\inbar{\,\vrule height1.5ex width.4pt depth0pt}
\def\IC{\relax\hbox{$\inbar\kern-.3em{\rm C}$}}
\def\IR{\relax{\rm I\kern-.18em R}}
\def\IP{\relax{\rm I\kern-.18em P}}

%
%

%
\catcode`\@=11
\def\slash#1{\mathord{\mathpalette\c@ncel{#1}}}
\overfullrule=0pt

\def\NN{{\cal N}}

\def\underrel#1\over#2{\mathrel{\mathop{\kern\z@#1}\limits_{#2}}}

\catcode`\@=12


%

\def\det{{\rm det}}

\def\det{{\rm det}}
\def\exp{{\rm exp}}


\lref\SakaiCN{
T.~Sakai and S.~Sugimoto,
``Low energy hadron physics in holographic QCD,''
Prog.\ Theor.\ Phys.\  {\bf 113}, 843 (2005)
[arXiv:hep-th/0412141].
}

\lref\SonET{
  D.~T.~Son and M.~A.~Stephanov,
  ``QCD and dimensional deconstruction,''
  Phys.\ Rev.\ D {\bf 69}, 065020 (2004)
  [arXiv:hep-ph/0304182].
}

\lref\BabingtonVM{
  J.~Babington, J.~Erdmenger, N.~J.~Evans, Z.~Guralnik and I.~Kirsch,
  ``Chiral symmetry breaking and pions in non-supersymmetric gauge /  gravity
  duals,''
  Phys.\ Rev.\ D {\bf 69}, 066007 (2004)
  [arXiv:hep-th/0306018].
}
\lref\EvansIA{
  N.~J.~Evans and J.~P.~Shock,
  ``Chiral dynamics from AdS space,''
  Phys.\ Rev.\ D {\bf 70}, 046002 (2004)
  [arXiv:hep-th/0403279].
}
\lref\GhorokuSP{
  K.~Ghoroku and M.~Yahiro,
  ``Chiral symmetry breaking driven by dilaton,''
  Phys.\ Lett.\ B {\bf 604}, 235 (2004)
  [arXiv:hep-th/0408040].
}

\lref\ErlichQH{
  J.~Erlich, E.~Katz, D.~T.~Son and M.~A.~Stephanov,
  ``QCD and a holographic model of hadrons,''
  Phys.\ Rev.\ Lett.\  {\bf 95}, 261602 (2005)
  [arXiv:hep-ph/0501128].
}

\lref\KarchPV{
  A.~Karch, E.~Katz, D.~T.~Son and M.~A.~Stephanov,
  ``Linear confinement and AdS/QCD,''
  arXiv:hep-ph/0602229.
}

\lref\AHJK{
  E.~Antonyan, J.~A.~Harvey, S.~Jensen and D.~Kutasov,
  ``NJL and QCD from string theory,''
  arXiv:hep-th/0604017.
}

\lref\WittenZW{
  E.~Witten,
  ``Anti-de Sitter space, thermal phase transition, and confinement in  gauge
  theories,''
  Adv.\ Theor.\ Math.\ Phys.\  {\bf 2}, 505 (1998)
  [arXiv:hep-th/9803131].
}

\lref\HawkingDH{
  S.~W.~Hawking and D.~N.~Page,
  ``Thermodynamics Of Black Holes In Anti-De Sitter Space,''
  Commun.\ Math.\ Phys.\  {\bf 87}, 577 (1983).
}

\lref\NambuTP{Y.~Nambu and G.~Jona-Lasinio,
``Dynamical Model Of Elementary Particles Based On An Analogy With
Superconductivity. I,''Phys.\ Rev.\  {\bf 122}, 345 (1961).
}

\lref\HatsudaPI{
T.~Hatsuda and T.~Kunihiro,
``QCD phenomenology based on a chiral effective Lagrangian,''
Phys.\ Rept.\  {\bf 247}, 221 (1994)
[arXiv:hep-ph/9401310].
}

\lref\VolkovKW{
M.~K.~Volkov and A.~E.~Radzhabov,
``Forty-fifth anniversary of the Nambu-Jona-Lasinio model,''
arXiv:hep-ph/0508263.
}

\lref\KlevanskyQE{
  S.~P.~Klevansky,
  ``The Nambu-Jona-Lasinio model of quantum chromodynamics,''
  Rev.\ Mod.\ Phys.\  {\bf 64}, 649 (1992).
}

\lref\BuballaQV{
  M.~Buballa,
  ``NJL model analysis of quark matter at large density,''
  Phys.\ Rept.\  {\bf 407}, 205 (2005)
  [arXiv:hep-ph/0402234].
}
\lref\GrossJV{
D.~J.~Gross and A.~Neveu,
``Dynamical Symmetry Breaking In Asymptotically Free Field Theories,''
Phys.\ Rev.\ D {\bf 10}, 3235 (1974).
}

\lref\GiveonSR{
A.~Giveon and D.~Kutasov,
``Brane dynamics and gauge theory,''
Rev.\ Mod.\ Phys.\  {\bf 71}, 983 (1999)
[arXiv:hep-th/9802067].
}

\lref\SakaiYT{
  T.~Sakai and S.~Sugimoto,
  ``More on a holographic dual of QCD,''
  Prog.\ Theor.\ Phys.\  {\bf 114}, 1083 (2006)
  [arXiv:hep-th/0507073].
}

\lref\ItzhakiDD{
N.~Itzhaki, J.~M.~Maldacena, J.~Sonnenschein and S.~Yankielowicz,
``Supergravity and the large N limit of theories with sixteen
supercharges,''
Phys.\ Rev.\ D {\bf 58}, 046004 (1998)
[arXiv:hep-th/9802042].
}

\lref\KruczenskiUQ{
  M.~Kruczenski, D.~Mateos, R.~C.~Myers and D.~J.~Winters,
  ``Towards a holographic dual of large-N(c) QCD,''
  JHEP {\bf 0405}, 041 (2004)
  [arXiv:hep-th/0311270].
}

\lref\KarchSH{
  A.~Karch and E.~Katz,
  ``Adding flavor to AdS/CFT,''
  JHEP {\bf 0206}, 043 (2002)
  [arXiv:hep-th/0205236].
}

\lref\StephanovWX{
  M.~A.~Stephanov,
  ``QCD phase diagram and the critical point,''
  Prog.\ Theor.\ Phys.\ Suppl.\  {\bf 153}, 139 (2004)
  [Int.\ J.\ Mod.\ Phys.\ A {\bf 20}, 4387 (2005)]
  [arXiv:hep-ph/0402115].
}

\lref\ASY{
  O.~Aharony, J.~Sonnenschein and S.~Yankielowicz,
  ``A holographic model of deconfinement and chiral symmetry restoration,''
  arXiv:hep-th/0604161.
}

\lref\HorigomeXU{
  N.~Horigome and Y.~Tanii,
  ``Holographic chiral phase transition with chemical potential,''
  arXiv:hep-th/0608198.
}

\lref\ChamblinTK{
  A.~Chamblin, R.~Emparan, C.~V.~Johnson and R.~C.~Myers,
  ``Charged AdS black holes and catastrophic holography,''
  Phys.\ Rev.\ D {\bf 60}, 064018 (1999)
  [arXiv:hep-th/9902170].
}
\lref\Olver{
F.~W.~J.~Olver, ``Asymptotics and special functions'', 
A~K~Peters,Wellesley,1997
}

\lref\ShuryakSE{
  E.~V.~Shuryak,
  ``Strongly coupled quark-gluon plasma: The status report,''
  arXiv:hep-ph/0608177.
}

\lref\ParnachevDN{
  A.~Parnachev and D.~A.~Sahakyan,
  ``Chiral phase transition from string theory,''
  Phys.\ Rev.\ Lett.\  {\bf 97}, 111601 (2006)
  [arXiv:hep-th/0604173].
}

\lref\AharonyDA{
  O.~Aharony, J.~Sonnenschein and S.~Yankielowicz,
  ``A holographic model of deconfinement and chiral symmetry restoration,''
  arXiv:hep-th/0604161.
}

\lref\BenincasaEI{
  P.~Benincasa and A.~Buchel,
  ``Hydrodynamics of Sakai-Sugimoto model in the quenched approximation,''
  Phys.\ Lett.\ B {\bf 640}, 108 (2006)
  [arXiv:hep-th/0605076].
}

\lref\AntonyanQY{
  E.~Antonyan, J.~A.~Harvey and D.~Kutasov,
  ``The Gross-Neveu model from string theory,''
  arXiv:hep-th/0608149.
}

\lref\AntonyanPG{
  E.~Antonyan, J.~A.~Harvey and D.~Kutasov,
  ``Chiral symmetry breaking from intersecting D-branes,''
  arXiv:hep-th/0608177.
}

\lref\BasuEB{
  A.~Basu and A.~Maharana,
   ``Generalized Gross-Neveu models and chiral symmetry breaking from string
  theory,''
  arXiv:hep-th/0610087.
}

\lref\CaronHuotTE{
  S.~Caron-Huot, P.~Kovtun, G.~D.~Moore, A.~Starinets and L.~G.~Yaffe,
  ``Photon and dilepton production in supersymmetric Yang-Mills plasma,''
  arXiv:hep-th/0607237.
}

\lref\SonSD{
  D.~T.~Son and A.~O.~Starinets,
   ``Minkowski-space correlators in AdS/CFT correspondence: Recipe and
  applications,''
  JHEP {\bf 0209}, 042 (2002)
  [arXiv:hep-th/0205051].
}

\lref\KarchSH{
  A.~Karch and E.~Katz,
  ``Adding flavor to AdS/CFT,''
  JHEP {\bf 0206}, 043 (2002)
  [arXiv:hep-th/0205236].
}

\lref\GepnerQY{
  D.~Gepner and S.~Sekahr Pal,
  ``Chiral symmetry breaking and restoration from holography,''
  arXiv:hep-th/0608229.
}

\lref\KimGP{
  K.~Y.~Kim, S.~J.~Sin and I.~Zahed,
  ``Dense hadronic matter in holographic QCD,''
  arXiv:hep-th/0608046.
}

\lref\MateosNU{
  D.~Mateos, R.~C.~Myers and R.~M.~Thomson,
  ``Holographic phase transitions with fundamental matter,''
  Phys.\ Rev.\ Lett.\  {\bf 97}, 091601 (2006)
  [arXiv:hep-th/0605046].
}

\lref\MateosYD{
  D.~Mateos, R.~C.~Myers and R.~M.~Thomson,
  ``Holographic Viscosity of Fundamental Matter,''
  arXiv:hep-th/0610184.
}

\lref\PeetersIU{
  K.~Peeters, J.~Sonnenschein and M.~Zamaklar,
   ``Holographic melting and related properties of mesons in a quark gluon
  plasma,''
  arXiv:hep-th/0606195.
}

\Title{\vbox{\baselineskip12pt
\hbox{hep-th/0610247}
}}
{\vbox{\centerline{Photoemission with Chemical Potential}
       \centerline{from QCD Gravity Dual}
\vskip.06in
}}
\centerline{Andrei Parnachev${}^a$ and David A. Sahakyan${}^b$}
\bigskip
\centerline{{\it ${}^a$C.N.Yang Institute for Theoretical Physics, SUNY, Stony Brook, NY 11794-3840, USA}}
\centerline{{\it ${}^b$Department of Physics, Rutgers University, Piscataway, NJ 08854-8019, USA }}
 \vskip.1in \vskip.1in \centerline{\bf Abstract}  
\noindent
We consider a $D4-D8-\bar D8$ brane construction 
which gives rise to a large $N$ QCD at sufficiently small energies.
Using the gravity dual of this system, we study chiral phase transition 
at finite chemical potential and temperature and find a line
of first order phase transitions in the phase plane.
We compute the spectral function and the photon emission rate.
The trace of the spectral function is monotonic at vanishing chemical
potential, but develops some interesting features as
the value of the chemical potential is increased.
\vfill

\Date{November 2006}
   

\newsec{Introduction}

There are indications that heavy ion collisions at RHIC
produce strongly coupled quark gluon plasma (sQGP)
(see \ShuryakSE\ for a recent review and a list of references).
String theory gives rise to a number of gauge/gravity
dual pairs where strongly coupled gauge theory is related to
gravity.
Hence, one may expect gauge/gravity duality (at finite temperature)
to be particularly useful for studying the qualitative properties of sQGP.
Moreover, one may hope that gauge theory intuition and direct
connection with experiment will bring new insights about
gravity itself.

A particularly attractive model where the low energy
degrees of freedom are just that of large $N_c$ QCD arises
from the collection of $N_c$ D4 and $N_f$ $D8-\bar D8$ branes
\SakaiCN.
(For some work in this direction see 
\refs{\AHJK\ParnachevDN\AharonyDA\BenincasaEI\PeetersIU\AntonyanQY
\AntonyanPG\KimGP\HorigomeXU\GepnerQY-\BasuEB}  )
In this model one of the spatial directions along the  $D4$ branes 
is compactified on a circle of radius $R_4$, which sets the scale
of the glueball masses and confinement-deconfinement transition
in the theory.
Another important parameter in the theory is asymptotic separation 
of the $D8$ and $\bar D8$ branes, denoted by $L$.
It sets the scale of the meson masses and determines the temperature
of chiral phase transition in the theory \refs{\ParnachevDN\AharonyDA}.
In the regime where $L\ll R_4$ the two sets of phenomena
happen at widely separated scales \AHJK.

In this paper we restrict the analysis to this regime and  
extend the results of \ParnachevDN\ to nonzero chemical 
potential $\mu$, dual to baryon number. 
We find the line of first order phase transitions in the $T-\mu$ 
plane.
We also compute the spectral function and the photoemission 
spectrum in the quark-gluon plasma
and find an intriguing picture.
While at small chemical potential the spectral function
is featureless, at larger values of $\mu$ it exhibits a characteristic
peak at intermediate frequencies. We also observe that the photoemission spectrum is redshifted as we increase the value of $\mu$.
The nature of these phenomena is not completely clear to us,
and it remains to be seen whether these are generic features
or they are associated with the particular string model\foot{
The emission of photons associated with the weakly gauged $U(1)_R$ symmetry
of $\NN=4$ SYM has been previously calculated in \CaronHuotTE.
As is well known, in this case nonzero chemical potential 
dual to the $U(1)_R$ charge destabilizes the theory.}.

The rest of the paper is organized as follows.
In the next Section we review the near-horizon geometry of
the D4 branes at finite temperature and zero chemical potential.
As usual, in the limit $g_s N_c\gg 1$, $N_c\rightarrow\infty$, $N_f=finite$,
one can use the DBI action of $D8$ branes to study the physics of fundamental matter.
At sufficiently high temperature or chemical potential 
the thermodynamically preferred
configuration is the one where $D8$ branes fall into the horizon.
In this situation baryonic $U(1)$ is no longer broken and the corresponding
current can be coupled to a photon via the term of the form $e J_\mu A^\mu$.
The number of photons emitted per unit volume per unit time is given by
\eqn\defg{ d\Gamma={d^3k\over (2\pi)^3} {e^2\over2|k|}
                    \eta^{\mu\nu} C_{\mu\nu}^< (K) |_{k^0=|k|}   }
where
\eqn\cmunu{  C_{\mu\nu}^<(K)\equiv 
 \int d^4 X e^{-i K X} \langle J_\mu(0) J_\nu(X)\rangle=
  n_b(k^0) \chi_{\mu\nu}(K)    }
In eq. \cmunu\ $n_b(k^0)=1/(e^{\beta k^0}-1)$ and
$\chi_{\mu\nu}(K)$ is the spectral function, proportional to the imaginary 
part of the retarded current-current correlation function,
\eqn\chiret{  \chi_{\mu\nu}(K)=-2 Im C^{ret}_{\mu\nu}(K)  }
which we compute using Lorenzian  AdS/CFT 
prescription of \SonSD.

In Section 3 we explain how the phase transition takes
place at finite chemical potential.
We generalize the photoemission computation for the case
of nonvanishing chemical potential in Section 4.
Surprisingly, we find that the spectral function
exhibits a bump whose physical significance is not clear to us.
We discuss our results in Section 5.
Asymptotic behavior of the spectral function is
computed in appendix.

\newsec{Photoemission at zero chemical potential} 

We will consider the near-horizon geometry of the D4-branes in the
limit studied in \AHJK, where the direction transverse to the D8 
branes is non-compact.
The near-horizon geometry of the $D4$-branes at finite temperature is: 
\eqn\metricss{ ds^2=\left(U\over R\right)^{3\over2}\left(dx_i dx^i+f(U)dt^2+(dx^4)^2\right)+
\left(U\over R\right)^{-{3\over2}}\left({dU^2\over f(U)}+U^2d\Omega_4^2\right)} 
where $f(U)=1-U_{T}^3/U^3$.
Here $t$ is the Euclidean time, $i=1,\ldots,3$; $U$ and $\Omega_4$ label the
radial and angular directions of $(x^5,\ldots,x^9)$. The parameter $R$ is given by
\eqn\rdefn{R^3 = \pi g_s N_c l_s^3=\pi\lambda}
where in the last equality and in the rest of the paper we set $\alpha'=1$.
The fourbrane geometry also has a non-trivial dilaton and RR-four form
\eqn\dfourdil{e^\Phi=g_s\left(U\over R\right)^{3\over4}\,,
\quad F_4=dC_3={2\pi N_c\over V_S}\epsilon_4}
where $V_S$ and $\epsilon_4$ are the volume and the volume form of the unit $4$-sphere.   
Finite temperature implies that $t$ is a periodic variable, 
\eqn\tper{ t\sim t+\beta  }
On the other hand, in order for \metricss\ to describe a non-singular space,
$t$ must satisfy
\eqn\xfourpd{ t=t+{4\pi R^{3/2}\over 3 U_{T}^{1/2}}   }
Hence, the temperature is related to the minimal value of $U$,
denoted by $U_T$, as
$T=3 U_{T}^{1/2}/4\pi R^{3/2}$.

The induced metric on the D8 branes takes the following form (we assume that the temperature is sufficiently high, so the chiral symmetry is restored \ParnachevDN)
\eqn\indmet{
 ds^2=\left({U\over R}\right)^{3/2}
  \left(\delta_{ij}dx^i dx^j-f(U)dt^2\right)
    +\left({U\over R}\right)^{-3/2}{dU^2\over f(U)}
    +R^{3/2} U^{1/2} d\Omega_4^2   }

In order to find the photon emission rate we need to find current-current correlators.
According to gauge/gravity duality the current-current correlators are computed
by analyzing the linearized perturbations of the dual gauge field, which in our case is the elecromagnetic field living on the $D8$ brane. In the following we compute photoemission due to a single brane. The total result contains an additional factor of $2N_f$.   

At non-zero temperature, rotation plus the gauge invariance implies that the 
correlator has the form
\eqn\corrform{C_{\mu\nu}^{ret}(k_\mu)=P^T_{\mu\nu}(k_\mu)
\Pi^T(k_\mu)+P^L_{\mu\nu}(k_\mu)\Pi^L(k_\mu)} 
where the non-zero components of the $P^T_{\mu\nu}(k_\mu)$ are
\eqn\PT{P^T_{ij}(k_\mu)=\delta_{ij}-{k_ik_j\over\vec k^2}}
and
\eqn\PL{P^L_{\mu\nu}=\eta_{\mu\nu}-{k_\mu k_\nu\over k^2}-P^T_{\mu\nu}(k_\mu)}
Substituting in \chiret\ we find that the trace of the spectral function is
\eqn\chirettrace{\chi^\mu_\mu=-4{Im}\Pi^T(k_\mu)-2{Im}\Pi^L(k_\mu)}
In this paper we are interested in computing the rate of production of  real photons, hence we will only consider lightlike momenta. 
In this case $\Pi^L(k_\mu)$ vanishes (otherwise the correlator \corrform\
would be singular on the lightcone) and only $\Pi^T(k_\mu)$ will contribute to the trace of the spectral function.   

Now we turn on electromagnetic field on the brane. 
We will be interested in field configurations which do not depend on the coordinates on $S^4$, we also set $A_M$ along the $S^4$ to zero. 
The DBI action 
is
\eqn\dbitu{  S=-g_s T_8 V_S R^3\int d^4 x\int dU U e^{-\Phi}\sqrt{-\det (g_{AB}+2\pi F_{AB})}}
%
where $(A,B=0,\cdots 3, U)$, $T_8=1/(2\pi)^8 g_s$ is the tension of the brane and $V_S=8\pi^2/3$ is the volume of the unit four-sphere. Expanding it in the second order in the field strength we find 
\eqn\dbiscnd{\eqalign{S=&{-}(2\pi^2)T_8 V_S R^{3/2}\int d^4 x\int dU U^{5/2}\left[{-}{1\over f(U)}\left({R\over U}\right)^3 \vec E^2{-}F_{0U}^2{+}\right.\cr
&\qquad\qquad\left.\left(R\over U\right)^3 \sum_{i<j}F_{ij}^2{+}f(U)\sum_{i}F_{iU}^2\right]}}
It is convenient to make a change of variables
\eqn\uy{  y={2 R^{3/2}\over\sqrt{U}},\qquad U={4 R^3\over y^2}  }
which brings \dbiscnd\ to the form 
\eqn\dbity{  S=-{64\over 3} \pi^4 T_8 R^{6}\int d^4 x
             \int dy y^{-2} \left[-{1\over f(y)}\vec E^2-F_{0y}^2+\sum_{i<j}F_{ij}^2+f(y)\sum_i F_{iy}^2 \right]   }
where 
\eqn\fofy{  f(y)=1-{y^6\over y_T^6},\qquad y_T={3\over2\pi T}       } 
The space is restricted to lie in the region $y\in(0,y_T)$, whose
upper limit corresponds to the black hole horizon.
It is convenient to perform Fourier transform in the space-time
\eqn\fouri{A_C(x^{\mu}, y)=\int {d^4 k\over (2\pi)^4} e^{ik_\mu x^\mu} A_C(k_\mu,y).}
Choosing $k_\mu=(-\omega,0,0,q)$, we find the following e.o.m. satisfied by the transverse electric field
\eqn\eqmo{E''_{\perp}+\left({f'\over f}-{2\over y}\right)E'_{\perp}+{\omega^2-q^2f\over f^2}E_{\perp}=0}
Let us discuss the asymptotics of this equation in the two regimes: $y\rightarrow y_T$ and $y\rightarrow 0$. 
In the near horizon limit the equation is solvable and $E_{\perp}$ is
\eqn\hor{E_{\perp}=(y_T-y)^{\pm iw/2}}
where $w=\omega y_T/3=\omega/2\pi T$. To compute the retarded correlators we should impose the incoming wave boundary condition, which fixes the sign in the exponent of the \hor\ to be negative.
Now let us discuss the opposite limit $y\rightarrow 0$. In this regime there are again two solutions which have the following asymptotics
\eqn\infsol{\eqalign{F_I&=1+a_3 y^3/y_T^3+\cdots\cr
F_{II}&=y^3/y_T^3+\cdots\cr
}}
Setting $a_3=0$ uniquely determines $F_I$. The solution of \eqmo\ with the incoming wave boundary condition has the form
\eqn\eqsol{E_{\perp}={\cal A} F_I+{\cal B} F_{II},}
where ${\cal B}$ is proportional to $\cal A$.

The boundary action for $E_{\perp}$ can be determined from
\eqn\bndact{{\delta S\over \delta E_{\perp}(y=0,k_\mu)}|_{EOM}=\lim_{y\rightarrow 0} {\delta S\over \delta 
E_{\perp}'(y,k_\mu)}=-{128\pi^4 T_8 R^6\over 3\omega^2} \lim_{y\rightarrow 0} y^{-2}f(y)E_{\perp}'~.}
Using \eqsol\ and varying this equation with respect to $E_{\perp}$, we find
\eqn\bnactwo{{\delta^2 S\over \delta A_{\perp}(0,k_{\mu})\delta A_{\perp}(0,-{k_{\mu}})}=\omega ^2{\delta^2 S\over \delta E_{\perp}(0,k_{\mu})\delta E_{\perp}(0,-{k_\mu})}=-{128\pi^4 T_8 R^6 y_T^{-3}} {{\cal B}\over {\cal A}}}
Hence 
\eqn\impt{{Im}(\Pi^T(k_{\mu}))=
-{4\pi\over 27} (\lambda T)T^2 N_c 
{Im}\left[{{\cal B}\over {\cal A}}\right]}
and the trace of the spectral function for lightlike momenta
\eqn\specfun{\chi_\mu^\mu=-4 Im\Pi^T(q=\omega)={16\pi\over 27}\lambda T^3 N_c {Im}\left[{\cal B\over \cal A}\right]}

To determine the spectral function we solve the equation \eqmo\ numerically with the incoming wave boundary condition and determine ${{\cal B}\over \cal A}$ as a function of $\omega$ (we set $q=\omega$). The result is depicted on Fig. 1.

For better numerical convergence near horizon it is convenient to rewrite the equation \eqmo\ in terms of slow varying mode defined as
\eqn\slowvar{E_{\perp}=(y_T-y)^{-{iw\over 2}}\varphi(y)~.}
Substituting in \eqmo\ we find that $\varphi$ satisfies
\eqn\eqmophi{\eqalign{
\varphi''(y)+&\left[{iw\over y_T-y}-{6y^5\over y_T^6-y^6}-{2\over y}\right]\varphi'(y)+\cr
&\left[\left(-{w^2\over 4}+{iw\over 2}\right){1\over (y_T-y)^2}-\left({6y^5\over y_T^6-y^6}+{2\over y}\right){iw\over 2}{1\over y_T-y}+
{9w^2y^6y_T^4\over (y_T^6-y^6)^2}\right]\varphi(y)=0
}
} 
As expected the exponents of this equation near $y=y_T$
are $0$ and $iw$. The solution with zero exponent 
corresponds to  $E_{\perp}$ with  incoming wave boundary condition. 

\midinsert\bigskip{\vbox{{\epsfxsize=3in
        \nobreak
    \centerline{\epsfbox{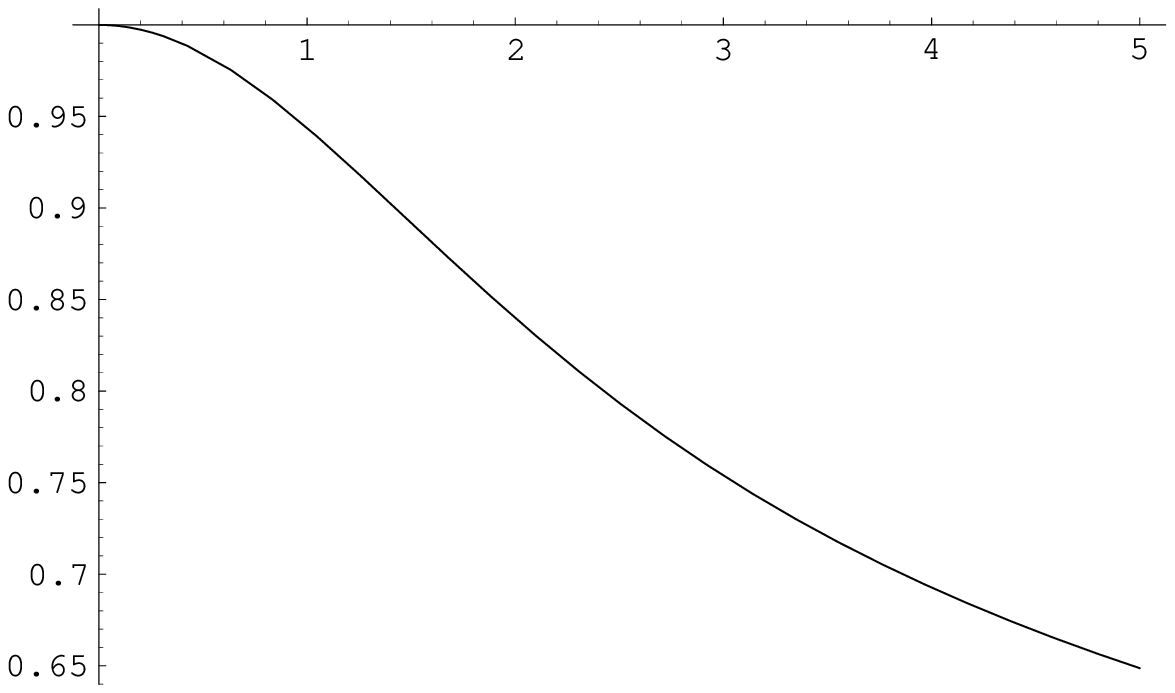}}
        \nobreak\bigskip
    {\raggedright\it \vbox{
{\bf Fig 1.}
{\it  $3\chi_\mu^\mu(\omega)/(\omega y_T)$ in the units of 
${16\pi\over 27}\lambda T^3 N_c$ as a function of $\omega y_T$.
}}}}}}
\bigskip\endinsert
Next we would like to discuss the asymptotic behavior of $\chi^\mu_\mu$ at high and low temperatures.
We start by discussing the low temperature regime $\omega y_T\rightarrow\infty$. It turns out that in
this regime we can solve $\eqmo$ analytically using generalized WKB  approximation. We perform this calculation in Appendix A and find
\eqn\specfunappr{\chi_\mu^\mu={16\pi\over 27}(\lambda T)T^2 N_c {Im}\left[{\cal B\over \cal A}\right]\simeq
{16\pi\over 81}\lambda T^3 N_c\sin\left({3\pi\over 8}\right){\Gamma\left({5\over 8}\right)\over 
\Gamma\left({11\over 8}\right)}\left({\omega y_T\over 8}\right)^{3\over 4}}
This approximation works very well at high frequencies as can be seen 
from Fig.2. 
\midinsert\bigskip{\vbox{{\epsfxsize=3in
        \nobreak
    \centerline{\epsfbox{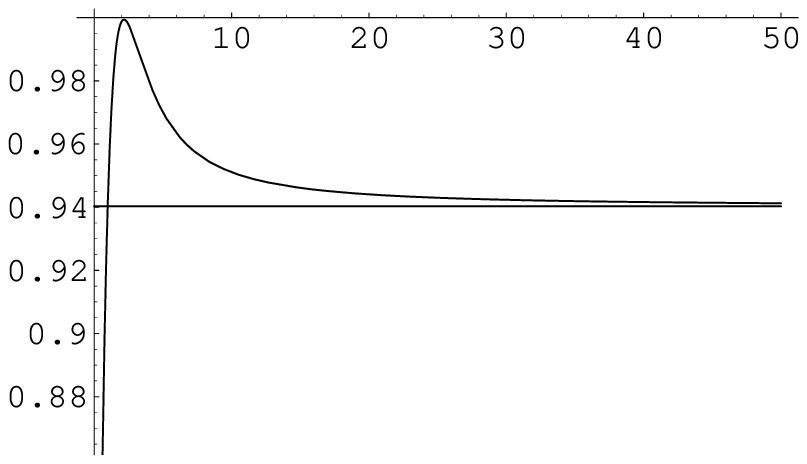}}
        \nobreak\bigskip
    {\raggedright\it \vbox{
{\bf Fig 2.}
{\it  $3\chi_\mu^\mu(\omega)/(\omega y_T)^{3/4}$ in the units of 
${16\pi\over 27}\lambda T^3 N_c$
as a function of $\omega y_T$. The constant line represents the approximate solution  \specfunappr.
}}}}}}
\bigskip\endinsert
\noindent
In principle one can improve this result by including higher order terms in the expansion in powers of $\omega^{-1}$.

In the high temperature regime we can solve the equation \eqmo\ perturbatively.
The leading behavior is computed  appendix A and yields
\eqn\specfunhigh{\chi_\mu^\mu={16\pi\over 27}(\lambda T)T^2 N_c {Im}\left[{\cal B\over \cal A}\right]\simeq
{16\pi\over 81}\lambda T^3 N_c\omega y_T~.}

\bigskip
\newsec{Chiral phase transition at finite chemical potential}

In this section we Wick rotate time coordinate to study the chiral 
phase transition at finite chemical potential.
Turning on chemical potential is equivalent to turning on non-trivial, purely imaginary $A_0^{(0)}(U)$.  Then the average $4d$ charge density is given by the  variation of the action w.r.t. $A_0^{(0)}(\infty)$ 
\eqn\variat{\rho=iT{\delta S\over \delta A_0^{(0)}(U=\infty)}|_{EOM}=iT\lim_{U\rightarrow\infty}{\delta S\over \delta A_0'^{(0)}(U)}
}
To find the action we allow non-trivial dependence $x^4\equiv\tau$ on the radial coordinate $U$ and compute the induced metric on the world volume of the D-brane. The resulting action is\foot{We use $A_U=0$ gauge. }
\eqn\dbitut{  S={T_8 V_S R^{3/2}\over T}\int d^3 x
               \int dU U^{5/2}\sqrt{1+4\pi^2{F_{0U}^{(0)}}^2+(U/R)^3 f(U)\tau'^2}   } 
It is convenient to replace $A_0^{(0)}\rightarrow i A_0^{(0)}$, so that
$\mu=A_0^{(0)}(\infty)+const$ (we will determine this constant below). 
Then the equations of motion which follow from \dbitut\ are
\eqn\eqmotion{\eqalign{
{d\over dU}\left[{2\pi U^{5/2} A_0'^{(0)}\over\sqrt{1-(2\pi A_0'^{(0)})^2+\left({U\over R}\right)^3f(U)\tau'^2}}\right]&=0\cr
{d\over dU}\left[{U^{11/2} f(U)\tau'\over\sqrt{1-(2\pi A_0'^{(0)})^2+\left({U\over R}\right)^3f(U)\tau'^2}}\right]&=0
}} 
which should be solved with boundary conditions $\tau(\infty)=\pm L/2$ and $A_0^{(0)}(\tau=\pm L/2)=\mu+const$.
As in the case of zero chemical potential there are two types of solutions:
$\tau'=0$, which has $U(N_f)\times U(N_f)$ chiral symmetry and the U shaped
solution, which breaks the symmetry down to $U(N_f)$. The phase in which chiral symmetry is broken does not have free charges, since the quarks form $q\bar q$ pairs which are neutral (the baryons are infinitely heavy in $N_c\rightarrow\infty$ limit). 
 Hence we expect that this phase should be unaffected by non-zero chemical potential. On the other hand  the phase in which the chiral symmetry is restored has free charges and should be sensitive to non-zero chemical potential. We  come to the same conclusion by analyzing the solutions of \eqmotion. Indeed it is easy to see that for the U shaped solution \eqmotion\ imply that $A_0^{(0)}$ is a monotonic function of $\tau$, hence in order to satisfy the boundary conditions one has to have $A_0'^{(0)}=0$\foot{The equations \eqmotion\ were obtained in \HorigomeXU, but their  $A_0^{(0)}$
for the U shaped solution is not a continuous function at $\tau=0$.}.
The situation is very similar to the Hawking-Page transition in AdS space in     
the presence of chemical potential. In this case the thermal AdS solution does not allow non-zero charge density, while the AdS black hole does \ChamblinTK.
The reason for this behavior from the CFT side is the same as in our situation: the AdS solution corresponds to the confined phase in which we do not have free charges.

For $\tau'=0$ the action \dbitut\ takes the following form
\eqn\cehmpot{S={T_8V_S R^{3/2}\over T}\int d^3 x\int dU U^{5/2}\sqrt{1-4\pi^2 {F_{0U}^{(0)}}^2}}
 which yield e.o.m 
\eqn\eomchem{{2\pi U^{5/2}{A_0'}^{(0)}\over \sqrt {1-4\pi^2 
({A_0'}^{(0)})^2}}=C }
To fix the constant in the relation between $\mu$ and $A_0^{(0)}(\infty)$
we require $4d$ charge density to vanish for $\mu=0$. This implies
\eqn\chempotgauge{\mu=A_0^{(0)}(\infty)-A_0^{(0)}(U_T)}
Setting $A_0^{(0)}(U_T)=0$, the chemical potential is 
\eqn\chempot{\mu=A_0^{(0)}(\infty)={1\over 2\pi}\int_{U_T}^\infty dU\sqrt{C^2\over C^2+U^5}={C^{2/5}\over 2\pi}\left[{\Gamma({3\over 10})
\Gamma({6\over 5})\over\sqrt\pi}-{U_T\over C^{2/5}}\,_2F_1\left({1\over 5},{1\over 2},{6\over 5},-{U_T^5\over C^2}\right)\right]}
The thermal expectation value of $4d$ charge density is related to $C$ as follows
\eqn\chrgden{\rho=2\pi T_8 V_S R^{3/2} C~.}
The relation between the chemical potential and the charge density simplifies considerably for $C\gg U_T^{5/2}$ and $C\ll U_T^{5/2}$.
For the former we find
\eqn\romu{\rho\simeq {\sqrt 2\over 12 \pi^2}\left({\Gamma({3\over 10})
\Gamma({6\over 5})\over\sqrt\pi}\right)^{-5/2}N_c\lambda^{-1/2}\mu^{5/2}} 
while for the latter
\eqn\roT{\rho\simeq N_c\left[ {4\pi\over 27}(\lambda T)T^2\mu+
{3^5\over 2^7 13\pi^3}{\mu^3\over (\lambda T)}\right]~.}
Rewriting the conditions of reliability for  \romu\ and \roT\ in terms of $\mu$,
$\lambda$ and $T$, we find that \romu\ is reliable for $\mu\gg(\lambda T)T$, while \roT\ for $\mu\ll(\lambda T)T$.
It is instructive to compare these relations with corresponding relations for $4d$ free fermions. In this case the charge density is
\eqn\freeferm{
\rho_{free}={N_c\over 6\pi^2} \left[\mu \pi^2 T^2+\mu^3\right]
} 
We see that while small $\mu$ behavior in our system bears some similarity with the free fermion case,  the large $\mu$ behavior is completely different.  

To understand the details of the chiral symmetry restoration we need to compare the grand potentials of the two phases
\eqn\diff{\eqalign{
\delta\Omega=\Omega_{st}-\Omega_U&=T_8V_SR^{3/2}\left[\int_{U_T}^{U_0} dU 
{U^5\over\sqrt{U^5+C^5}}+\right.\cr
&\left.\int_{U_0}^\infty dU U^{5/2}\left[{U^{5/2}\over\sqrt{U^5+C^5}}-\sqrt{1+{f(U_0)U_0^8\over f(U)U^8-f(U_0)U_0^8}}\right]\right]
}}
where $U_0$ is the value of $U$ at the tip of the U shaped solution.
Let us first discuss the $LT\rightarrow 0$ case. Solving numerically the equation $\delta \Omega=0$ we find the critical value for $C$
\eqn\num{{U_0\over C_c^{2/5}}\simeq 1.83} 
Using \eqmotion\ to express $U_0$ in terms of the asymptotical separation $L$ and substituting into \chempot\ we find
\eqn\chemcrit{\mu_c\simeq {8R^3L^{-2}\over 1.83}{\Gamma(3/10)\Gamma(6/5)\over\sqrt\pi}\left[{\Gamma(9/16)\over
\Gamma(6/5)}\right]^2\simeq 0.22 L^{-2}\lambda}
The phase diagram at non-zero $T$ is given by
\eqn\phase{\mu_c(T)=\phi(LT)\lambda L^{-2},}
where $\phi(x)$ is monotonically decreasing function, which can be determined from \diff.  
 
\bigskip

\newsec{Photoemission at finite chemical potential}
Now we would like to compute the photoemission in the presence of non-zero chemical  potential. As before we turn on electromagnetic field on the brane and expand \dbitu\ in the second order in the field strength. 
Note that $\det(g_{AB}+2\pi {\cal F}_{AB})$ is an even function of $\cal F$, hence 
\eqn\detexp{\det(g_{AB}+2\pi {\cal F}_{AB})=\det g_{AB}(1+({\cal F}^2- terms)+({\cal F}^4- terms))}
We set ${\cal A}_0=A_0^{(0)}+A_0$, ${\cal A}_B=A_B$ for $B\ne 0$,
${\cal F}_{0U}=F_{0U}^{(0)}+F_{0U}$ and 
${{\cal F}_{AB}}=F_{AB}$ for $AB\ne 0U$.
In addition to the DBI action there is also contribution coming from the CS term
\eqn\CSaction{S_{CS}=i\mu_9\int{\rm Tr}\left[\exp(2\pi {\cal F}_2)\wedge C_3\right]=
i\mu_9 {(2\pi)^3\over 3!}\int {\cal F}\wedge {\cal F}\wedge {\cal F}\wedge C_3}
where $\mu_9=1/(2\pi)^8 g_s$.
Integrating this action over the internal $S^4$ we find 
\eqn\CSint{S_{CS}= i {N_c\over 24\pi^2}\int {\cal A}\wedge {\cal F}\wedge {\cal F}}
Using the expansion \detexp\ we find\foot{Note that there is a term in action linear in $F_{0U}$. This term enforces the equations of motion for $F_{0U}^{(0)}$ hence it is omitted in the expression for the action. }
\eqn\dbichem{\eqalign{S=&-2\pi^2T_8 V_S R^{3/2}\int d^4 x\int {dU\over \sqrt{U^5+C^2}}\left[-{U^5+C^2\over f(U)}\left({R\over U}\right)^3 \vec E^2-{(U^5+C^2)^2\over U^5}F_{0U}^2+\right.\cr
&\left.\left({R\over U}\right)^3 U^5\sum_{i<j}F_{ij}^2+(U^5+C^2)f(U)\sum_i F_{iU}^2\right]+S_{CS}\cr
 } }
The action in coordinates \uy\ is
\eqn\dbichemy{\eqalign{S=&-{64\over 3}\pi^4T_8 R^6\int d^4 x\int {dy y^{-2}\over \sqrt{1+\tilde C^2 y^{10}/y_T^{10}}}\left[-{1+\tilde C^2 y^{10}/y_T^{10}\over f(y)}\vec E^2-{(1+\tilde C^2 y^{10}/y_T^{10})^2}F_{0y}^2+\right.\cr
&\left.\sum_{i<j}F_{ij}^2+(1+\tilde C^2 y^{10}/y_T^{10})f(y)\sum_i F_{iy}^2\right]+S_{CS}\cr
 } }
where
\eqn\tc{\tilde C^2={C^2y_T^{10}\over 2^{10}R^{15}}\sim {\rho^2\over 
(\lambda T)^4 N_c^2T^6}~.}   
The variation of the Chern-Simons term is
\eqn\CSeom{\delta S= i {N_c\over 8\pi^2} \int \delta A\wedge F^{(0)}\wedge F} 
Taking into account that
\eqn\FO{{dA_0^{(0)}\over dy}=-8R^3{\tilde Cy^2y_T^{-5}\over 2\pi\sqrt{1+\tilde C^2y^{10}/y_T^{10}}} }
we find that the equations of motion for the transverse modes are
\eqn\eqmot{\eqalign{
E''_1&+\left({f'\over f}+{3\over y}-{5\over y}{1\over (1+\tilde C^2y^{10}/y_T^{10})}\right)E'_1+{\omega^2-{q^2\over (1+\tilde C^2 y^{10}/y_T^{10})}f\over f^2}E_1+\cr
&\qquad\qquad\qquad\qquad\qquad\qquad\qquad\qquad\qquad\quad 3{q\tilde C y^4y_T^{-5}\over (1+\tilde C^2y^{10}/y_T^{10})f}E_2=0\cr
E''_2&+\left({f'\over f}+{3\over y}-{5\over y}{1\over (1+\tilde C^2y^{10}/y_T^{10})}\right)E'_2+{\omega^2-{q^2\over (1+\tilde C^2 y^{10}/y_T^{10})}f\over f^2}E_2-\cr
&\qquad\qquad\qquad\qquad\qquad\qquad\qquad\qquad\qquad\quad
3{q\tilde C y^4y_T^{-5}\over (1+\tilde C^2y^{10}/y_T^{10})f}E_1=0}}
Note that the asymptotics of this equation as $y\rightarrow y_T$ and $y\rightarrow 0$ are the same as for \eqmo. 
The solution with boundary conditions \hor\ for $E_\bot=E_1$
and $E_2|_{y=y_T}=E_2'|_{y=y_T}=0$ has a form
\eqn\hora{  E_1 = {\cal A}_1 F_I + {\cal B}_1 F_{II}   }
\eqn\horb{  E_2 = {\cal A}_2 F_I + {\cal B}_2 F_{II}   }
where the coefficients ${\cal A}_i$, ${\cal B}_i$
can be computed numerically.
By rotational invariance, boundary conditions \hor\ for $E_\bot=E_2$
and $E_1|_{y=y_T}=E_1'|_{y=y_T}=0$ give rise to the solution
\eqn\horaa{  E_2 = {\cal A}_1 F_I + {\cal B}_1 F_{II}   }
\eqn\horbb{  E_1 = -{\cal A}_2 F_I -{\cal B}_2 F_{II}   }
To compute $\chi^\mu_\mu$, we need to take the linear combination
of the solutions such that the leading term in $E_2$ (or $E_1$) near the
boundary vanishes, and then repeat \bndact\--\impt.
 The spectral function for lightlike momenta is
\eqn\specfunchem{\chi_\mu^\mu=
-4 Im\Pi^T(q=\omega)=
{16\pi\over 27}(\lambda T)T^2 N_c {Im}
\left[{{\cal A}_1{\cal B}_1+{\cal A}_2{\cal B}_2
\over {\cal A}_1^2+{\cal A}_2^2}\right]}
Solving \eqmot\ with incoming wave boundary condition numerically we find the spectral function in the case of non-zero chemical potential. The result is depicted on Fig 3.  
\midinsert\bigskip{\vbox{{\epsfxsize=3in
        \nobreak
    \centerline{\epsfbox{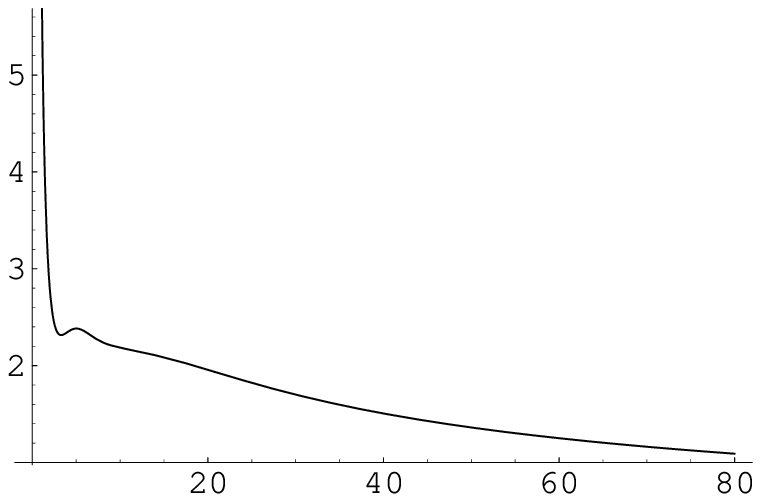}}
        \nobreak\bigskip
    {\raggedright\it \vbox{
{\bf Fig 3.}
{\it  $3\chi_\mu^\mu(\omega)/(\omega y_T)$ in the units of 
${16\pi\over 27}\lambda T^3 N_c$ as a function of $\omega y_T$ for $\tilde C=200$.
}}}}}}
\bigskip\endinsert
\noindent
The result is somewhat surprising: while the spectral function for zero chemical potential is featureless, we see that at non-zero chemical potential things change and we get a second maximum. The nature of this peak is not clear to us.

The leading behavior of the spectral function at high temperature is computed  in appendix A 
\eqn\specfunhigh{\chi_\mu^\mu={16\pi\over 27}\lambda T^3 N_c  {Im}
\left[{{\cal A}_1{\cal B}_1+{\cal A}_2{\cal B}_2
\over {\cal A}_1^2+{\cal A}_2^2}\right]\simeq
{16\pi\over 27}(\lambda T)T^2 N_c\omega y_T\sqrt{1+\tilde C^2}~.}

\bigskip
\newsec{Discussion}

In this paper we extended the results of \ParnachevDN\ to the
case of nonvanishing chemical potential.
We found a line of first order phase transitions in 
the $\mu-T$ plane which intersects the axes at $(\mu=0,T=0.15 L^{-1})$
and $(\mu_c=0.22 L^{-2}\lambda,T=0)$.
Note that the value of $\mu_c$ is of the order of constituent quark mass
in the phase with broken chiral symmetry\foot{ The constituent quark mass is given
by the mass of the string which goes from the tip of D8 brane all the way
down to the horizon. 
In this system it is much larger then the meson mass.}.

We also studied photoemission at finite temperature 
and chemical potential. At small chemical potential the 
spectral function is featureless while at larger values of $\mu$ it exhibits a characteristic peak at intermediate frequencies. 

Using \defg\ we rewrite the photon emission rate as function of the frequency 
\eqn\defgg{{d\Gamma\over d\omega}={e^2\over 4\pi}
{\omega\chi_\mu^\mu\over e^{\omega/ T}-1} }
Then for the zero chemical potential we plot the spectrum on Fig 4. using the results of section 2.
For comparison we also plot the black body radiation spectrum on the same graph  
\eqn\deffpp{{d\Gamma_P\over d\omega}\propto
{\omega^2\over e^{\omega/ T}-1}}
\midinsert\bigskip{\vbox{{\epsfxsize=3in
        \nobreak
    \centerline{\epsfbox{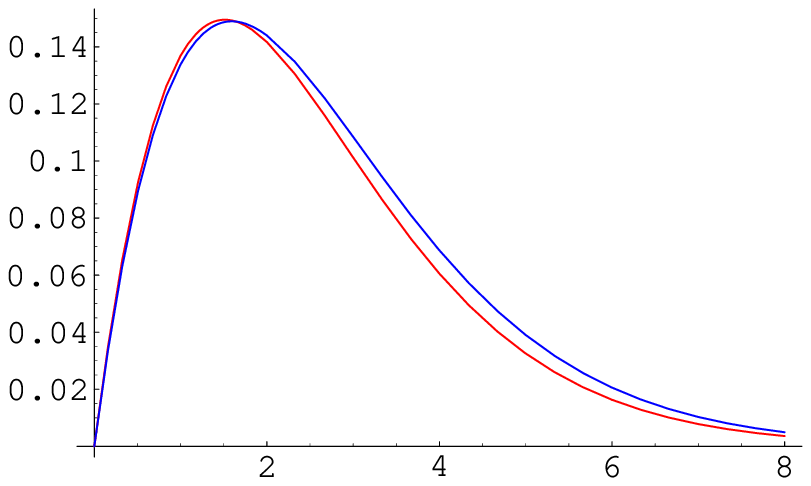}}
        \nobreak\bigskip
    {\raggedright\it \vbox{
{\bf Fig 4.}
{\it  $d\Gamma/d\omega$ in the units of 
${4e^2\over 81}(\lambda T)T^3 N_c$ as a function of $\omega/T$--red curve and
the black body radiation spectrum (in arbitrary units)--blue curve.
}}}}}}
\bigskip\endinsert
\noindent  
We observe that at zero chemical potential the spectrum of radiation is very similar to the Planckian spectrum. As we turn on  the chemical potential things start to change. The spectrum becomes more and more red-shifted as we increase the chemical potential. 
\midinsert\bigskip{\vbox{{\epsfxsize=3in
        \nobreak
    \centerline{\epsfbox{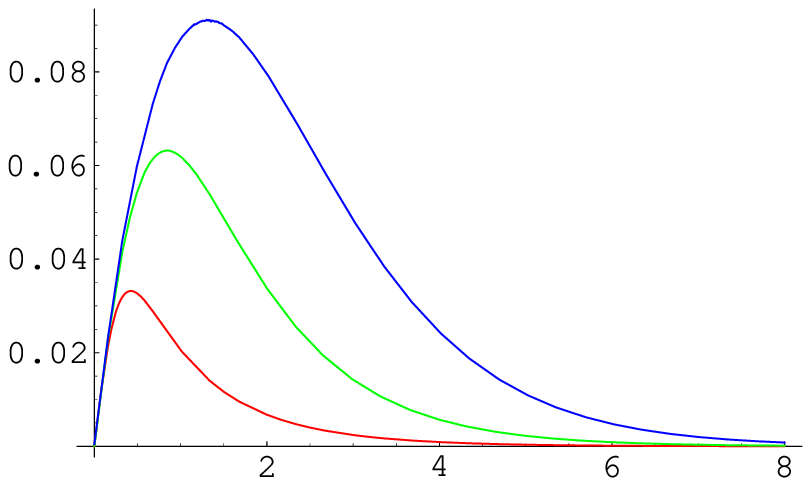}}
        \nobreak\bigskip
    {\raggedright\it \vbox{
{\bf Fig 5.}
{\it  ${1\over\sqrt{1+\tilde C^2}}{d\Gamma\over d\omega}$ in the units of 
${4e^2\over 81}(\lambda T)T^3 N_c$ as a function of $\omega/T$.
Red, green and blue curves correspond to $\tilde C=1000,\,100,\,10$ respectively.
}}}}}}
\bigskip\endinsert
\noindent  

The natural question is whether something similar happens
in other models.
One particularly simple way of incorporating fundamental
matter involves adding D7 branes to the stack of D3 branes \KarchSH.
The resulting theory is four dimensional, but
the presence of bosonic fields charged under the $U(1)_V$
does not allow turning on nonzero chemical potential\foot{One may 
still induce nonzero charge density by condensing the corresponding
bosonic fields.}.
Similar problems persist for the D4-D6 system recently studied in
\refs{\MateosNU\MateosYD}\foot{
The value of $\mu$ in this system is bounded from above by the 
scalar mass, while we observed that the nontrivial features in the spectral
density appear at very large values of $\mu$.}.

The model we consider does not have charged bosonic fields.
On the other hand one must bear in mind that it involves
four dimensional fundamental matter and five dimensional 
gauge field and only becomes four dimensional QCD for energies
much smaller then $1/R_4$.
This is the reason for the appearance of the factors $(\lambda T)$
in the expressions for the spectral functions in this paper.
(In comparison, such factors are absent in the four-dimensional
model of \CaronHuotTE.)
It would be nice to better understand the physical origin of the features of the photoemission spectrum in strongly coupled gauge theories
at nonvanishing chemical potential.

\bigskip
{\noindent\bf Acknowledgements:}

We would like to thank D. Mateos, E. Shuryak and I. Zahed for useful discussions.
We thank Aspen Center for Physics where this work was initiated.
This work was supported in part by  DOE grant DE-FIG02-96ER40949.

\appendix{A}{Asymptotics for the spectral function}

In this appendix we find the asymptotic behavior of the spectral function 
at high and low temperatures. 
We start by discussing the low temperature regime $\omega y_T\rightarrow\infty$. It turns out that in
this regime we can solve $\eqmo$ analytically using generalized WKB  approximation. We will follow the approach of \Olver.
First we introduce new variable 
\eqn\newfiled{\tilde E_{\perp}\equiv {\sqrt{1-y^6}\over y} E_{\perp}}  
to eliminate the first derivative term in \eqmo\foot{Here and below we rescaled $y\rightarrow y/y_T$ and $\omega\rightarrow \omega y_T$ for simplicity.}
\eqn\eqmonew{\tilde E_{\perp}''+\left[{\omega^2y^6\over (1-y^6)^2}-
{2y^{12}-13y^6+2\over y^2(1-y^6)^2}\right]\tilde E_{\perp}=0}
In the $\omega\rightarrow\infty$ regime the first term in the brackets dominates 
almost everywhere on the segment $y\in (0,1)$. As we approach $y=0$ point the second term starts contributing. Keeping only the most singular contribution from the second term we obtain the approximate equation
\eqn\eqmoneappr{\tilde E_{\perp}''+\left[{\omega^2y^6\over (1-y^6)^2}-
{2\over y^2}\right]\tilde E_{\perp}=0}
Rewriting \eqmoneappr\ in terms of new variables
\eqn\newvar{\eqalign{z^6&\equiv{\left(dy\over dz\right)}^2 {y^6\over (1-y^6)^2}\cr
W&\equiv{\tilde E_{\perp}\over\sqrt{\dot y}}
}}
we obtain
\eqn\eqmosim{{d^2 W\over dz^2}+\left[\omega^2 z^6-{2\over z^2}\right]W=0.}
This equation can be solved analytically in terms of Bessel functions 
\eqn\sol{W(z)={\cal A}\Gamma\left({5\over 8}\right)
\left({\omega\over 8}\right)^{{3\over 8}}
\sqrt{z}J_{-{3\over 8}}({\omega z^4/ 4)}+{\cal B}\Gamma\left({11\over 8}\right)
\left({\omega\over 8}\right)^{-{3\over 8}}\sqrt{z}J_{{3\over 8}}({\omega z^4/ 4)}}
As we will see shortly the first term corresponds to the $F_{I}$ in \eqsol, while the second term corresponds to $F_{II}$. Hence we identified the integration constants with $\cal A$ and $\cal B$.  
Then the approximate solution to \eqmo\ is
\eqn\apprsol{E_{\perp}(y))={z(y)^{3/2}\over y^{3/2}}y W(z(y))~.} 
To find the spectral function we will need the behavior of $z$ as 
$y\rightarrow 1$ and $y\rightarrow 0$. Using \newvar\ we find
\eqn\Y{\eqalign{z^4/4={1\over 2\sqrt 3}
&\left(\arctan\left[{1+2y\over \sqrt 3}\right]-
\arctan\left[{2y-1\over \sqrt 3}\right]\right)-{1\over 6}\log(1-y)-\cr
&{1\over 6}\log(1+y)+
{1\over 12}\log(1-y+y^2)+{1\over 12}\log(1+y+y^2)-{\pi\over 6\sqrt 3}}}
Hence 
\eqn\Yas{\eqalign{z=&y+\cdots,\quad{\rm for}\,\,\, y\rightarrow 0\cr
z^4=&-{2\over 3}\log(1-y)+\cdots,\quad{\rm for}\,\,\,y\rightarrow 1}
}
Using the asymptotic behavior of the Bessel functions 
\eqn\asymBess{J_\alpha(x)\simeq {1\over \Gamma(\alpha+1)}\left({x\over 2}\right)^\alpha\quad {\rm for}\,\,\,x\rightarrow 0}
we confirm the identifications made in \sol. To find the spectral function we should look at the solution \apprsol\ in the opposite regime $y\rightarrow 0$ ($z\rightarrow \infty$) and impose the incoming wave boundary conditions
\eqn\Ebehav{E_\perp\simeq (1-y)^{-i\omega/6}\simeq e^{i\omega z^4\over 4} }
Using the asymptotic form of Bessel functions at large value of the argument
\eqn\asymBesslarge{J_\alpha(x)\simeq \sqrt{2\over\pi x} \cos\left(x-{\alpha\pi\over 2}-{\pi\over 4}\right)}
 we find
\eqn\specfunapprA{{Im}\left[\cal B\over \cal A\right]=
\sin\left({3\pi\over 8}\right){\Gamma\left({5\over 8}\right)\over 
\Gamma\left({11\over 8}\right)}\left({\omega\over 8}\right)^{3\over 4}}
The case of the finite chemical potential can be treated similarly, although the calculation is more involved and we will not pursue it here. 

Now let us discuss the high temperature case $\omega\rightarrow 0$.  
In this regime we can solve equations \eqmo\ and the corresponding equation for the finite chemical potential \eqmot\ perturbatively in $\omega=3w$
\eqn\pert{E_{\perp}(y)=(1-y)^{-{iw\over 2}}\varphi(y)=
(1-y)^{-{iw\over 2}}(F^{(0)}(y)+w F^{(1)}(y)+\cdots)}
We start by discussing the zero chemical potential case. 
Then $\varphi(y)$ satisfies equation \eqmophi. In the $w^0$ order the equation has two solutions one of which is constant while the other is 
logarithmically divergent at the horizon. The solution with incoming wave boundary conditions correspond to the constant solution which we set to one. 
Substituting \pert\ into \eqmophi\ we find that $F^{(1)}(y)$ at $w$ order satisfies
\eqn\emoF{
F^{(1)}(y)''-\left[{6y^5\over 1-y^6}+{2\over y}\right]F^{(1)}(y)'+
{i\over 2}\left[{1\over (1-y)^2}-\left({6y^5\over 1-y^6}+{2\over y}\right){1\over 1-y}\right]=0
} 
Solving \emoF\ we find
\eqn\solemF{F^{(1)}={i\over 12}\left((5-2i C_1)\log(1-y)-(1+2iC_1)\log\left({1+y+y^2\over 1+y^3}\right)\right)} 
The coefficient in front of the first term should be zero to satisfy the boundary condition. Using this explicit solution we find that near horizon $E_{\perp}$ is
\eqn\Eperp{E_{\perp}(y)=1+iwy^3+\cdots}
Hence to the leading order in $w$ 
\eqn\highT{{Im}\left[{\cal B\over \cal A}\right]\simeq w.} 
In the case of finite chemical potential we have
\eqn\pertchem{\eqalign{
E_1&=(1-y)^{-{iw\over 2}}\varphi_1(y)=
(1-y)^{-{iw\over 2}}(F_1^{(0)}(y)+w F_1^{(1)}(y)+\cdots)\cr
E_2&=(1-y)^{-{iw\over 2}}\varphi_2(y)=
(1-y)^{-{iw\over 2}}(F_2^{(0)}(y)+w F_2^{(1)}(y)+\cdots)
}}
The incoming wave boundary conditions for $E_1$ and $E_2|_{y=y_T}=E_2'|_{y=y_T}=0$ imply that
\eqn\zeroorder{F_1^{(0)}(y)=1\quad {\rm and}\quad F_2^{(0)}(y)=0} 
Then in the next to leading order the two equations in \eqmot\ decouple and we find that $F_1^{(1)}$ satisfies the following equation    
\eqn\emoFchem{\eqalign{F_1^{(1)}(y)''-&\left[{6y^5\over 1-y^6}-{3\over y}+
{5\over y}{1\over 1+\tilde C^2y^{10}}\right]F_1^{(1)}(y)'+\cr
&{i\over 2}\left[{1\over (1-y)^2}-\left({6y^5\over 1-y^6}-
{3\over y}+{5\over y}{1\over 1+\tilde C^2y^{10}}\right){1\over 1-y}\right]=0}}
$F_2^{(1)}$ satisfies similar equation, which we will not need at this order.
Solving \emoFchem\ we find
\eqn\solemFchem{F_1^{(1)}(y)=C_1\int_0^y{y^2\over (1-y^6)\sqrt{1+\tilde C^2y^{10}}}dy+
i\log(1-y)}
The integral above is logarithmically divergent near $y=1$. Imposing the regularity condition we find $C_1$. Near $y=0$ we can again approximate the integral by dropping the square root in the denominator. Hence we find
\eqn\Eperpchem{\eqalign{E_1(y)=\,&1+iw\sqrt{1+\tilde C^2}y^3+\cdots~,\cr
E_2(y)=\,&w F_2^{(1)}(y)+\cdots~}}
and
\eqn\highTch{ {Im}
\left[{{\cal A}_1{\cal B}_1+{\cal A}_2{\cal B}_2
\over {\cal A}_1^2+{\cal A}_2^2}\right]={Im}\left[{\cal B}_1\over {\cal A}_1\right]\simeq w
\sqrt{1+\tilde C^2}.}

\listrefs
\end